\documentclass[aps,pra,preprint,groupedaddress,showpacs,longbibliography]{revtex4-1}
\usepackage{amsmath,amssymb}
\usepackage{graphicx}
\usepackage{bm,bbm}
\usepackage{color}

\begin{document}
	
	
	\title{Nonclassical and semiclassical para-Bose states}
	
	\author{C. Huerta Alderete}
	\affiliation{
		Instituto Nacional de Astrof\'{\i}sica, \'Optica y Electr\'onica, Calle Luis Enrique Erro No. 1, Santa Maria Tonantzintla, Puebla CP 72840, M\'exico
	}%
	\author{Liliana Villanueva Vergara}
\affiliation{
	Instituto Nacional de Astrof\'{\i}sica, \'Optica y Electr\'onica, Calle Luis Enrique Erro No. 1, Santa Maria Tonantzintla, Puebla CP 72840, M\'exico
}%
	\author{B. M. Rodr\'iguez-Lara}%
	\email{bmlara@itesm.mx}
	\affiliation{%
		Photonics and Mathematical Optics Group, Tecnol\'ogico de Monterrey, Monterrey 64849, M\'exico.
	}%
	\affiliation{
		Instituto Nacional de Astrof\'{\i}sica, \'Optica y Electr\'onica, Calle Luis Enrique Erro No. 1, Santa Maria Tonantzintla, Puebla CP 72840, M\'exico
	}%
	\date{\today}
	
	\begin{abstract}
	Motivated by the proposal to simulate para-Bose oscillators in a trapped-ion setup [C. Huerta Alderete and B. M. Rodr\'iguez-Lara, Phys. Rev. A \textbf{95}, 013820 (2017)], we introduce an overcomplete, nonorthogonal basis for para-Bose Hilbert spaces. 
	The states spanning these bases can be experimentally realized in the trapped-ion simulation via time evolution. 
	The para-Bose states show both nonclassical and semiclassical statistics on their Fock state distribution, asymmetric field quadrature variances, and do not minimize the uncertainty relation for the field quadratures. 
	These properties are analytically controlled by the para-Bose order and the evolution time; both parameters might be feasible for fine tuning in the trapped-ion quantum simulation.
	\end{abstract}
	
	\pacs{03.65.-w, 03.67.-a, 37.10.Ty, 42.50.-p, 42.50.Ct}
	
	\maketitle
	
	\section{Introduction}
	
	Paraparticles in quantum mechanics arise from the fact that the Heisenberg-Born-Jordan relation for the quantum harmonic oscillator is not uniquely determined by the equations of motion \cite{Wigner1950p711} and are specifically related to deformations induced by the reflection operator \cite{Yang1951p788} that appears in the Calogero-Vasiliev oscillator \cite{Calogero1969p2191,Vasiliev1991p1115}.
	While paraparticles are allowed in the formalism of quantum mechanics, they have not been experimentally discovered as fundamental particles in nature \cite{Greenberg1965pB1155,Baker2015p929}. 
	Recently, we have proposed a quantum simulation of driven even-order para-Bose oscillators in a trapped-ion setup \cite{HuertaAlderete2017p013820}, for example,
	\begin{eqnarray}
	\hat{H} = \omega \hat{n} + g \left( \hat{A}^{\dagger} + \hat{A} \right),
	\end{eqnarray} 
	where the para-Bose operators of order $p$ fulfill the following bilinear commutation relations \cite{Green1953p270}:
	\begin{eqnarray}
	\left[ \hat{A}, \hat{A}^{\dagger} \right] &=& 1 + (p-1) \hat{\Pi}, \nonumber \\
	\left\{ \hat{A}, \hat{A}^{\dagger} \right\} &=& 2 \hat{n} +  p, \nonumber  \\
	\left[\hat{n} , \hat{A}^{\dagger} \right] = \hat{A}^{\dagger}, &\quad&  \left[\hat{n} , \hat{A} \right] = - \hat{A}.
	\end{eqnarray}
	We have used the parity operator given by $\hat{\Pi} = e^{i \pi \hat{n}}$; the operators $\hat{n}$, $\hat{A}^{\dagger}$, and $\hat{A}$ are the number, creation, and annihilation operators of the oscillator and the order parameter $p$ is a positive integer, $p=1,2,3\ldots$.
	The standard boson algebra showing Bose-Einstein statistics is recovered with order $p=1$.
	The work on para-Bose particles is extensive and, for example, their coherent states are well understood \cite{Sharma1978p2089,Sharma1981p78}.
	
	Here, we are interested in the properties of a peculiar set of states provided by simple time evolution in a particular realization of our para-Bose oscillators simulation,
	\begin{eqnarray}
	\vert p; igt \rangle = e^{ i g t \left(\hat{A}^{\dagger} + \hat{A} \right)} \vert p, 0 \rangle.
	\end{eqnarray}
	These states look like Gilmore-Perelomov coherent states but, formally, only the standard boson case, $p=1$, yields Gilmore-Perelomov coherent states  \cite{Gilmore1972p391,Perelomov1972p222} due to the characteristics of the para-Bose algebra.
	Nevertheless, we will provide in the following an analytic expression for these states and show that they span an overcomplete, nonorthogonal basis for the corresponding para-Bose Hilbert space.
	Then, we will address their para-Bose Fock occupation distribution, and show that it follows sub-, super-, and Poissonian statistics, calculated through the Mandel Q parameter, depending on the coherent parameter look-alike and the para-Bose order. 
	Thus, our previously proposed quantum simulation allows the engineering of highly nonclassical and semiclassical para-Bose states under the restrictions imposed by experimental constraints such as dissipation. 	Furthermore, we will show that these states do not minimize the Robertson uncertainty relation for the para-Bose field quadratures, nor are their quadrature variances equal for para-Bose particles beyond standard bosons.
	Finally, we will close with a summary and conclusion.

\section{Definition, completeness and orthogonality}

We can write a generalization of the states provided by time evolution of the trapped-ion simulation in a Gilmore-Perelomov coherent state look-alike form,
\begin{eqnarray}
\vert p; \alpha \rangle = e^{\alpha \hat{A}^{\dagger} - \alpha^{\ast} \hat{A}} \vert p; 0\rangle,
\end{eqnarray}
where the parameter $\alpha$ is a complex number that parametrizes the states, and the state $\vert p , 0 \rangle$ is the para-Bose vacuum state of order $p$.
Formally, only standard bosons, $p=1$, are Gilmore-Perelomov coherent states \cite{Gilmore1972p391,Perelomov1972p222},
\begin{eqnarray}
e^{\alpha \hat{A}^{\dagger} - \alpha^{\ast} \hat{A}}~\hat{A} ~e^{-(\alpha \hat{A}^{\dagger} - \alpha^{\ast} \hat{A})} \vert p=1; 0\rangle = 0.
\end{eqnarray}
Nevertheless, we will show that these para-Bose states can be interesting on their own.
It is cumbersome but straightforward to calculate their para-Bose Fock state decomposition, 
	\begin{eqnarray}
	\vert p; \alpha \rangle &=& e^{-\frac{\vert \alpha \vert^{2}}{2}} \sum_{j=0}^{\infty}  \frac{( \sqrt{2} \alpha )^{2j}}{(2j)!} \sqrt{ j! \left( \frac{p}{2} \right)_{j} }  \nonumber \\
	&&\times ~_{1}F_{1}\left( \frac{1-p}{2} , j + \frac{1}{2}, \frac{\vert \alpha \vert^{2}}{2} \right) \vert p; 2 j \rangle  \nonumber \\
	&&+ e^{-\frac{\vert \alpha \vert^{2}}{2}} \sum_{j=0}^{\infty} \frac{ ( \sqrt{2} \alpha)^{2j+1}}{(2j+1)!} \sqrt{ j! \left( \frac{p}{2} \right)_{j+1} }  \nonumber \\
	&&\times ~_{1}F_{1}\left( \frac{1-p}{2}, j + \frac{3}{2}, \frac{\vert \alpha \vert^{2}}{2} \right) \vert p; 2 j+1 \rangle,
	\end{eqnarray}
where the notations $\left( x \right)_{j}$ and $_{1}F_{1}(a,b;x)$ stand for the Pochhammer symbol and the confluent hypergeometric function of the first kind, respectively.
As the second argument of the confluent hypergeometric function fulfills $j+1/2 >0$, we can be sure that all of the amplitudes are entire functions of $p$ and $\alpha$ that converge for all finite $\alpha$ \cite{Lebedev1965}.
Here, we can also use  the fact that ${}_{1}F_{1}\left(0, b; z \right) = 1$ for any given $b$ and $z$ to recover the standard coherent states for bosons proposed by Sudarshan \cite{Sudarshan1963p277} and Glauber \cite{Glauber1963p2766},
\begin{eqnarray}
\vert p=1; \alpha \rangle &=& e^{- \frac{\vert \alpha \vert^{2}}{2}} \sum_{j=0}^{\infty}  \frac{ \alpha ^{j}}{\sqrt{j!}}  \vert p=1; j \rangle.
\end{eqnarray}

It is possible to demonstrate that our para-Bose states resolve the identity, 
\begin{eqnarray}
\frac{1}{\pi} \int d^{2} \alpha \vert p; \alpha \rangle \langle p; \alpha \vert  = \mathbbm{1}.
\end{eqnarray}
This can be accomplished by substitution of the para-Bose Fock state decomposition presented above, 
\begin{eqnarray}
\int d^{2} \alpha \vert p; \alpha \rangle \langle p; \alpha \vert &=& \pi \left[\sum_{j}  \frac{2^{2j+1} j! \left(\frac{p}{2}\right)_{j}}{\left[(2j)!\right]^{2}}  I_{even} \vert 2j \rangle \langle 2 j \vert \right. \nonumber \\
&& \left. +  \sum_{j}  \frac{2^{2j+2} j! \left(\frac{p}{2}\right)_{j+1}}{\left[(2j+1)!\right]^{2}}  I_{odd} \vert 2j + 1 \rangle \langle 2 j +1 \vert \right],
\end{eqnarray}
and reduced to prove the following integrals; 
\begin{eqnarray}
I_{even} &=& \int_{0}^{\infty} dr ~r^{4j+1} e^{-r^2} ~_{1}F^{2}_{1}\left(\frac{1-p}{2}, j+\frac{1}{2},\frac{r^{2}}{2}\right) = \frac{\left[(2j)!\right]^2}{2^{2j+1}j!\left(\frac{p}{2}\right)_{j}}, \\
I_{odd} &=& \int_{0}^{\infty} dr ~r^{4j+3} e^{-r^2} ~_{1}F^{2}_{1}\left(\frac{1-p}{2}, j+\frac{3}{2},\frac{r^{2}}{2}\right) =  \frac{\left[(2j+1)!\right]^2}{2^{2j+2}j!\left(\frac{p}{2}\right)_{j+1}}
\end{eqnarray}
which can also be computed numerically.
Here, orthogonality arises from the phase of the complex parameter $\alpha$, and the radial-like variable is its module, $ r = \vert \alpha \vert$.
Thus, the set of states $ \left\{ \vert p, \alpha \rangle \right \}$ with $p=1,2,3,\ldots$ and $\alpha \in \mathbb{C}$ form an overcomplete basis for the corresponding para-Bose Hilbert space of order $p$.

The basis is also non-orthogonal with overlaps that decay with the addition of the complex parameters moduli, in a similar fashion to standard coherent states,
\begin{eqnarray}
\langle p; \alpha \vert p; \beta \rangle &=& e^{-\frac{1}{2} \left( \vert \alpha \vert^2 + \vert \beta \vert^2 \right)}  \sum_{j} \left[ \frac{ (\alpha^{\ast} \beta )^{2j}}{(2j)!} A_{j}(\alpha, \beta) + \frac{ (\alpha^{\ast} \beta )^{2j+1}}{(2j+1)!} A_{j+1}(\alpha, \beta) \right]
\end{eqnarray}
where the auxiliary confluent hypergeometric product converges for finite parameters,
\begin{eqnarray}
A_{j}(\alpha,\beta) &=& \frac{\left(\frac{p}{2}\right)_{j}}{\left(\frac{1}{2}\right)_{j}} ~_{1}F_{1}\left(  \frac{1 -p}{2}, j + \frac{1}{2}, \frac{\vert \alpha \vert^{2}}{2} \right)  ~_{1}F_{1}\left(  \frac{1 -p}{2}, j + \frac{1}{2}, \frac{\vert \beta \vert^{2}}{2} \right) .
\end{eqnarray}
In short, these experimentally feasible states are the standard coherent states for just $p=1$ and  provide us with an overcomplete, nonorthogonal basis for the para-Bose Hilbert space of the order of $p \ge 2$.

\section{Para-bose Fock state distribution}

\begin{figure}[htbp]
\centering
\includegraphics[scale=1]{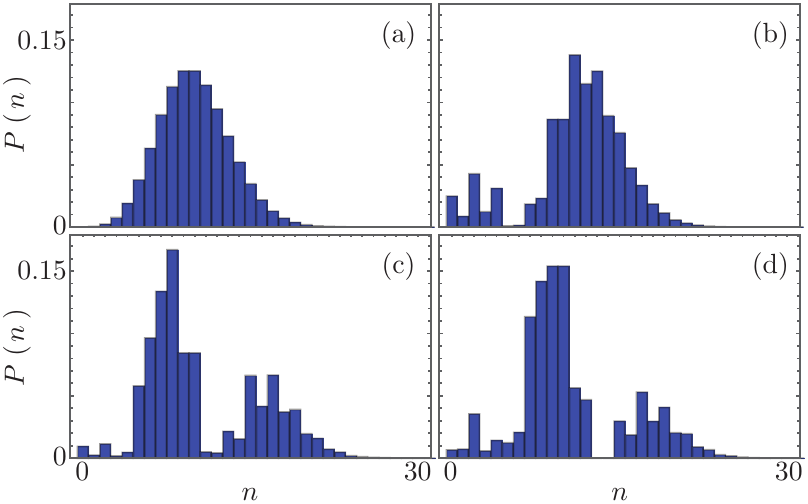}
\caption{Para-Bose Fock number state occupation distribution for some members $\vert p; \alpha \rangle$ of our overcomplete, nonorthogonal para-Bose basis with order (a) $p=1$, (b) $p=2$, (c) $p=5$, and (d) $p=6$, and complex parameter $\alpha = \sqrt{10}$. }
\label{fig:Fig1}
\end{figure}

Our closed form for the overcomplete, nonorthogonal para-Bose bases allows the study of their properties; for example, their para-Bose Fock number state occupation distribution,
\begin{eqnarray}
P(n) &=& \vert \langle p; n \vert p; \alpha \rangle \vert^2, \nonumber \\
	 &=& 	2^{n} \left( \frac{ \vert \alpha \vert^{n}}{n!} \right)^{2} e^{- \vert \alpha \vert^{2}} \nonumber \\
	&&\times \left\{ \begin{array}{ll} 
    \left( \frac{n}{2}\right)! ~ \left( \frac{p}{2}\right)_{\frac{n}{2}}~ \left\vert _{1}F_{1}\left( \frac{1-p}{2}, \frac{n+1}{2}, \frac{\vert \alpha \vert^2}{2}\right) \right\vert^2 , & n ~\mathrm{even}, \\
	 \left( \frac{n-1}{2}\right)! ~ \left( \frac{p}{2}\right)_{\frac{n+1}{2}} ~ \left\vert _{1}F_{1}\left( \frac{1-p}{2}, \frac{n+2}{2}, \frac{\vert \alpha \vert^2}{2} \right) \right\vert^2, & n ~\mathrm{odd}.
	\end{array} \right. \nonumber \\
	\end{eqnarray}
As expected, we recover the standard coherent states photon distribution for $p=1$ [Fig. \ref{fig:Fig1}(a)] and more complex distributions for higher orders; for example, for order $p=2$ [Fig. \ref{fig:Fig1}(b)], $p=5$ [Fig. \ref{fig:Fig1}(c)], and $p=6$ [Fig. \ref{fig:Fig1}(d)]. 
The effect of having para-Bose particles is such that distributions for contiguous orders have significant variations with respect to each other: Figs. \ref{fig:Fig1}(a) and \ref{fig:Fig1}(b) and Figs. \ref{fig:Fig1}(c) and \ref{fig:Fig1}(d).
	
The mean para-Bose excitation number, 
\begin{eqnarray}
\langle \hat{n} \rangle &=& \left\{ \begin{array}{ll} 
\vert \alpha \vert^{2} ~f(\vert\alpha\vert), & p = 2,  \\
\vert \alpha \vert^{2} + \frac{(p-1)}{2(p-2)} ~g(p,\vert\alpha\vert), & 
p \neq 2,
\end{array} \right.  
\end{eqnarray}
can also be given in terms of integer functions,  
\begin{eqnarray}
f(\vert\alpha\vert) &=& 1 + ~_{2}F_{2}\left(1,1; \frac{3}{2},2; -2 \vert \alpha \vert^2\right), \nonumber \\
g(p,\vert\alpha\vert) &=& 1- e^{ -2 \vert \alpha \vert^{2}} {}_{1}F_{1}\left(\frac{3-p}{2},\frac{1}{2}, 2\vert \alpha \vert^{2}\right),
\end{eqnarray}
involving the confluent hypergeometric function $_{1}F_{1}(a, b; x)$ and the generalized hypergeometric function $_{2}F_{2}(a_{1}, a_{2}; b_{1}, b_{2}; x)$ \cite{Lebedev1965}.
Note that the asymptotic values for the hypergeometric functions, 
\begin{eqnarray}
\lim_{\vert \alpha \vert \rightarrow \infty} ~_{2}F_{2}\left(1,1; \frac{3}{2},2; -2 \vert \alpha \vert^2\right) &=& 0, \nonumber \\ 
\lim_{\vert \alpha \vert \rightarrow \infty} e^{ -2 \vert \alpha \vert^{2}} {}_{1}F_{1}\left(\frac{3-p}{2},\frac{1}{2}, 2\vert \alpha \vert^{2}\right) &=& 0,  \quad p>2,
\end{eqnarray}
allow us to write the mean para-Bose excitation number for a large complex parameter module,
 \begin{eqnarray}
 \langle \hat{n} \rangle &\approx& \left\{ \begin{array}{ll} 
 \vert \alpha \vert^{2} + \frac{1}{4}\left[\ln(2 \vert \alpha\vert^2) -\psi\left(\frac{1}{2}\right)\right] , & p = 2,  \\
 \vert \alpha \vert^{2} + \frac{(p-1)}{2(p-2)}, & 
 p > 2, 
 \end{array} \right.  \qquad \vert \alpha \vert \gg 1, \nonumber \\
 \end{eqnarray}
where the function $\psi(x)$ is the digamma function \cite{Olver2010}. 
Note that in the limit of large para-Bose order and large complex parameter module, we can approximate the mean para-Bose excitation number by $ \langle \hat{n} \rangle \approx \vert \alpha \vert^{2} + 1/2$ for $p \gg 1$ and $\vert \alpha \vert \gg 1$. 
These asymptotic behaviors were confirmed numerically.

It is also possible to provide the variance for the mean para-Bose excitation number,
\begin{eqnarray}
\sigma^{2}_{n} =  \langle \hat{n}^{2} \rangle - \langle \hat{n} \rangle^{2},
\end{eqnarray}
but we leave it to the Appendix due to its size.
It is shorter to write its asymptotic behavior,
\begin{eqnarray}
\hspace{-.5cm}\sigma^{2}_{n} &\approx& \left\{ \begin{array}{ll} 
 \vert \alpha \vert^{2} \left[\frac{1}{2} \ln \left(2 \vert \alpha \vert^{2}\right) + \frac{3}{2} + \frac{1}{2}\psi \left(\frac{1}{2}\right)-\psi \left(\frac{3}{2}\right)\right] - \frac{1}{16}\psi^2 \left(\frac{1}{2}\right) \\ 
 - \frac{1}{8} \left[3 + \psi \left(\frac{1}{2}\right)+ \frac{1}{8} \ln\left(2 \vert\alpha\vert^{2}\right)\right] \ln \left(2 \vert \alpha \vert^{2}\right) + \frac{1}{4} + \frac{3}{8} \psi \left(\frac{1}{2}\right) , & p = 2,  \\
\frac{5}{2} \vert\alpha\vert^{2} + \frac{9}{8} \left[\ln \left(2 \vert\alpha\vert^{2}\right) - \psi \left(\frac{1}{2}\right)\right] - \frac{33}{16} , & p = 4,  \\
\vert \alpha \vert^{2}\frac{2p-3}{p-2} + \frac{p(p-1)}{4(p-4)(p-2)}, & p \neq 2,4,
\end{array} \right.   \vert \alpha \vert \gg 1,\nonumber \\
\end{eqnarray}
and we can realize that our para-Bose states with large complex parameter modulus always have a variance of the excitation number larger than its mean: $\sigma^{2}_{n} > \langle \hat{n} \rangle$ for  $p \ge 2$ and  $\vert \alpha \vert \gg 1$.

\begin{figure}[htbp]
		\centering
		\includegraphics[scale=1]{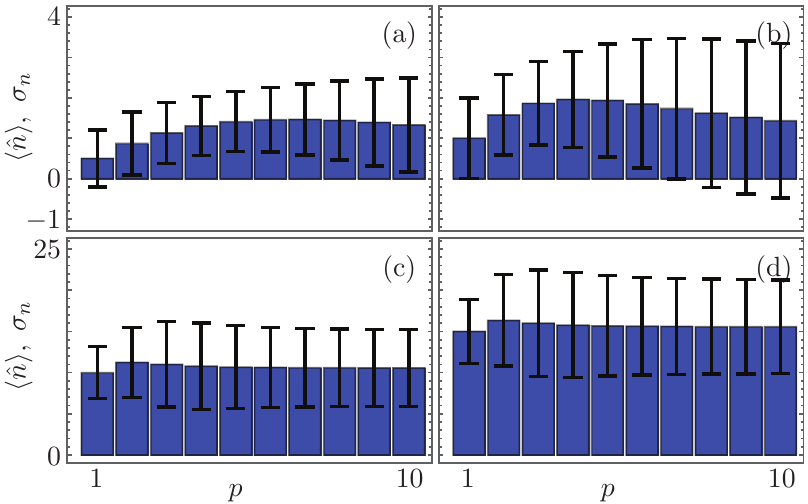}
		\caption{Mean para-Bose excitation number, $\langle \hat{n} \rangle$ in bars, and its standard deviation, $\sigma_{n}$ in error bars, for some members $\vert p; \alpha \rangle $ of our overcomplete, nonorthogonal para-Bose basis with complex parameter (a) $\alpha = \sqrt{1/2}$, (b) $\alpha = 1$, (c) $\alpha = \sqrt{10}$, and (d) $\alpha = \sqrt{15}$. }
		\label{fig:Fig2}
	\end{figure}

A more complex behavior emerges outside the asymptotic limit, as shown in Fig. \ref{fig:Fig2} for different coherent parameter values, $\alpha = \sqrt{1/2}$ [Fig. \ref{fig:Fig2}(a)], $\alpha = 1$ [Fig. \ref{fig:Fig2}(b)], $\alpha = \sqrt{10}$ [Fig. \ref{fig:Fig2}(c)], and $\alpha = \sqrt{15}$ [Fig. \ref{fig:Fig2}(d)], where we show the standard deviation as error bars, $\sigma_{n}$.
From the numerics outside the asymptotic regime, we can conjecture the following: 
First, for a given para-Bose order larger than two, there exists a family of isomodular complex parameters that makes the mean value of the para-Bose excitation number and its variance equal. 
For these complex parameters, lower- (larger-)order para-Bose states will have a mean excitation number that is larger (smaller) than its variance.
Second, there exists an isomodular family of states, $\vert \alpha \vert \sim 1.9018801$, that makes the mean value of the para-Bose excitation number and its variance equal for para-Bose order two, $p=2$. 
This value serves as the critical value after which the mean value of the excitation number will be smaller than its variance for any given para-Bose order.
Obviously, for any given complex parameter and standard bosons, the mean value of the excitation number and its variance are always equal, $\langle \hat{n} \rangle = \sigma^{2}_{n}= \vert \alpha \vert^{2}$ for $p=1$.
	
This peculiar behavior for the mean value of the para-Bose excitation number and its variance means our states will show sub-Poissonian, Poissonian, and super-Poissonian statistics depending on the para-Bose order and the complex parameter modulus. 
We can see this with the Mandel Q parameter \cite{Mandel1979p205},
\begin{eqnarray}
Q = \frac{\sigma^{2}_{n} - \langle \hat{n} \rangle}{\langle \hat{n} \rangle},
\end{eqnarray}
which is no longer a constant for our para-Bose states, again we provide the asymptotic for a large complex parameter modulus for the sake of space,
\begin{eqnarray}
\hspace{-1cm} Q &\approx& \left\{ \begin{array}{ll} 
\frac{4 \vert\alpha\vert^{4} + 4 \vert\alpha\vert^{2} \left[\ln \left(2 \vert\alpha\vert^2\right) + 3 - \psi \left(\frac{3}{2}\right)\right] + 2}{4 \vert\alpha\vert^{2} + \ln \left(2 \vert\alpha\vert^{2}\right) - \psi \left(\frac{1}{2}\right)} \\
-\vert\alpha\vert^2 - \frac{1}{4} \left[\ln \left(2 \vert\alpha\vert^2\right) - \psi \left(\frac{1}{2}\right)\right] - \frac{5}{2}, & p = 2,  \\
\frac{6}{4 \vert\alpha\vert^{2}+3} \left\{\vert\alpha\vert^{2} + \frac{3}{4} \left[\ln \left(2 \vert\alpha\vert^{2}\right) - \frac{5}{2} - \psi \left(\frac{1}{2}\right)\right]\right\} , & p = 4,  \\
\frac{(p-1) \Big[4\vert\alpha\vert^{2} (p-4) (p-2) - p^{2} + 11 p - 16 \Big]}{2 (p-4) (p-2) \Big[2 \vert\alpha\vert^{2} (p-2) + p-1 \Big]}, & 
p \neq 2,4.
\end{array} \right.   \vert \alpha \vert \gg 1, \nonumber \\
\end{eqnarray}
Upon close inspection, we can realize that in the limit of the large complex parameter modulus, the para-Bose Fock state distribution of our states is always super-Poissonian as expected, $Q > 0$ for $\vert \alpha \vert \gg 1$.

\begin{figure}[htbp]
	\centering
	\includegraphics[scale=1]{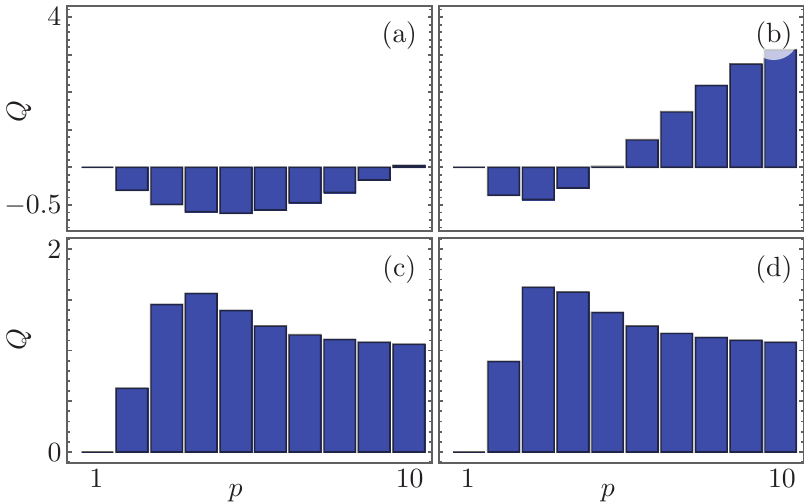}
	\caption{Mandel Q parameter for some members $\vert p; \alpha \rangle $ of our overcomplete, nonorthogonal para-Bose basis with complex parameter (a) $\alpha = \sqrt{1/2}$, (b) $\alpha = 1$, (c) $\alpha = \sqrt{10}$, and (d) $\alpha = \sqrt{15}$. }
	\label{fig:Fig3}
\end{figure}

The full expression for the Mandel Q parameter, which can be found in the Appendix, allows us to rewrite the conjectures mentioned above: For a given para-Bose order larger than two, there exists a family of isomodular complex parameters that yields Poissonian statistics for the excitation number distribution; lower- (larger-)order para-Bose states that belong to this isomodular complex parameter family will have sub- (super)-Poissonian statistics, shown in Figs. \ref{fig:Fig3}(a) and \ref{fig:Fig3}(b).
There exists a critical family of isomodular complex parameters, $\vert \alpha \vert \sim 1.9018801$, after which the excitation number distributions for para-Bose order $p\ge2$ follow super-Poissonian statistics and may be described with a semi-classical theory; see Figs. \ref{fig:Fig3}(c) and \ref{fig:Fig3}(d). 

\section{Para-Bose field quadratures and their uncertainty relation}
	
\begin{figure}[htbp]
	\centering
	\includegraphics[scale=1]{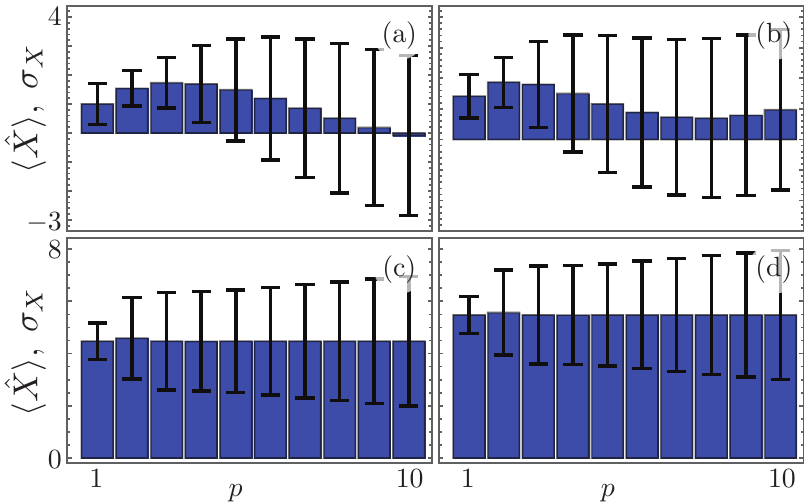}
	\caption{Expectation value of $\hat{X}$ quadrature, $\langle \hat{X} \rangle$ in bars, and its standard deviation, $\sigma_{X}$ in error bars, for some members $\vert p; \alpha \rangle $ of our overcomplete, nonorthogonal para-Bose basis with complex parameter (a) $\alpha = \sqrt{1/2}$, (b) $\alpha = 1$, (c) $\alpha = \sqrt{10}$, and (d) $\alpha = \sqrt{15}$. }
	\label{fig:Fig4}
\end{figure}

The mean values for the field quadratures,
\begin{eqnarray}
\hat{X} &=& \frac{1}{\sqrt{2}}  \left( \hat{A}^{\dagger} + \hat{A} \right) , \nonumber  \\
\hat{Y} &=& \frac{i}{\sqrt{2}}  \left( \hat{A}^{\dagger} - \hat{A} \right) ,
\end{eqnarray}
 of our overcomplete, nonorthogonal para-Bose basis are also integer functions, 	
\begin{eqnarray}
\langle \hat{X} \rangle &=& \frac{1}{\sqrt{2}} \left(\alpha + \alpha^{\ast}\right) h(p,\vert\alpha\vert), \nonumber \\
\langle \hat{Y} \rangle &=& \frac{i}{\sqrt{2}} \left(\alpha^{\ast} -  \alpha\right) h(p,\vert\alpha\vert). 
\end{eqnarray}
given in terms of confluent hypergeometric  functions of the first kind,
\begin{eqnarray}
h(p,\vert\alpha\vert) = 1 + (p-1) ~e^{-2\vert \alpha\vert^{2}} ~_{1}F_{1} \left( \frac{3-p}{2}, \frac{3}{2}, 2 \vert \alpha \vert^{2} \right).
\end{eqnarray}
This auxiliary function takes a unit value for both standard bosons, $h(1, \alpha)=1$, and large complex parameters, $h(p,\alpha) \approx 1$ for $\vert \alpha \vert \gg 1$, and we recover the standard boson value for the para-Bose field quadratures.
The second momenta of the para-Bose field quadratures can be found in the Appendix and used to construct the quadratures variances,
	\begin{eqnarray}
	\sigma^{2}_{X} &=& \left\{ \begin{array}{ll} 
	1 + \frac{1}{2} \left(\alpha + \alpha^{\ast}\right)^{2} \Big[f(\vert\alpha\vert) - h^{2}(2,\vert\alpha\vert)\Big] , & p = 2,  \\
	\frac{p}{2} + \frac{1}{2}\left(\alpha + \alpha^{\ast}\right)^{2} \left[1 + \frac{(p-1)}{2 \vert \alpha \vert^{2} (p-2)} g(p,\vert\alpha\vert) - h^{2}(p,\vert\alpha\vert)\right], & p \neq 2,
	\end{array} \right. \nonumber \\  
	\sigma^{2}_{Y} &=& \left\{ \begin{array}{ll} 
	1 - \frac{1}{2} \left(\alpha^{\ast} - \alpha\right)^{2} \Big[f(\vert\alpha\vert) - h^{2}(2,\vert\alpha\vert)\Big] , & p = 2,  \\
	\frac{p}{2} - \frac{1}{2}\left(\alpha^{\ast} - \alpha\right)^{2} \left[1 + \frac{(p-1)}{2 \vert \alpha \vert^{2} (p-2)} g(p,\vert\alpha\vert) - h^{2}(p,\vert\alpha\vert)\right], & p \neq 2,
	\end{array} \right.  \nonumber \\
	\end{eqnarray}
Figure \ref{fig:Fig4} show the mean value of the $\hat{X}$ para-Bose field quadrature as bars and its standard deviation $\sigma_{X}$ as error bars for para-Bose orders $p \in [1,10]$ and complex parameter values identical to previous figures $\alpha = \sqrt{1/2}$ [Fig. \ref{fig:Fig4}(a)], $\alpha = 1$ [Fig. \ref{fig:Fig4}(b)], $\alpha = \sqrt{10}$ [Fig. \ref{fig:Fig4}(c)], and $\alpha = \sqrt{15}$ [Fig. \ref{fig:Fig4}(d)]. 
While it is possible to provide asymptotic approximations for the variances of the field quadratures, which can be found in the Appendix, they portray no intelligible information beyond the behavior shown in the plots. 

\begin{figure}[htbp]
	\centering
	\includegraphics[scale=1]{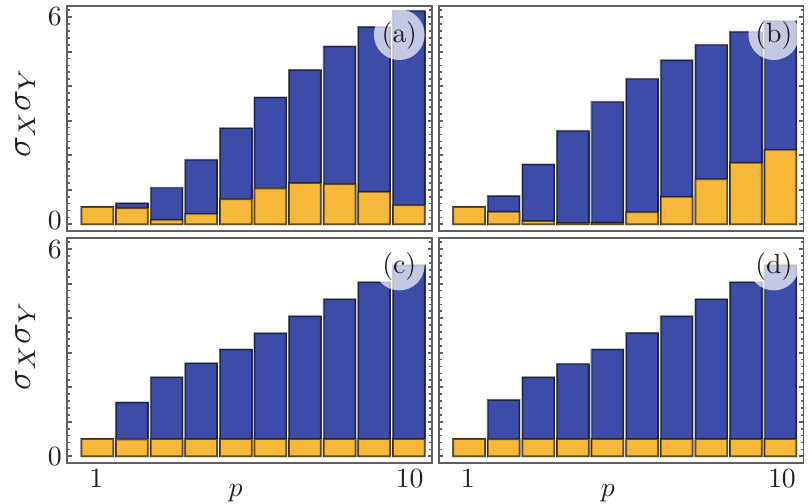}
	\caption{(Color online) Quadratures uncertainty relation, $\sigma_{X} \sigma_{Y}$, for some members $\vert p; \alpha \rangle $ of our overcomplete, nonorthogonal para-Bose basis with complex parameter (a) $\alpha = \sqrt{1/2}$, (b) $\alpha = 1$, (c) $\alpha = \sqrt{10}$, and (d) $\alpha = \sqrt{15}$ in blue (dark gray) bars and the lower bound for the corresponding Robertson uncertainty relation in gold (light gray) bars. }
	\label{fig:Fig5}
\end{figure}

We can also calculate the lower bound of the Robertson uncertainty relation for the field quadratures \cite{Robertson1929p163}, 
\begin{eqnarray}
\sigma_{X} \sigma_{Y} \ge \frac{1}{2} \left\vert 1 + \left( p-1\right) \langle \hat{\Pi} \rangle \right\vert,
\end{eqnarray}
which yields a simple closed form expression,
\begin{eqnarray}
\sigma_{X} \sigma_{Y} \ge \frac{1}{2} \left\vert 1 + \left( p-1\right)   e^{- 2 \vert \alpha \vert ^2} {}_{1}F_{1}\left( \frac{1-p}{2},\frac{1}{2}, 2 \vert \alpha \vert^{2} \right) \right\vert, \nonumber \\
\end{eqnarray}
which, again, converges to the standard boson expression for large para-Bose complex parameter values,
\begin{eqnarray}
\sigma_{X} \sigma_{Y} \ge \frac{1}{2}, \quad \vert \alpha \vert \gg 1, p\ge 2, 
\end{eqnarray}
 due to the asymptotic properties of the confluent hypergeometric function, 
\begin{eqnarray}
\lim_{\vert \alpha \vert \rightarrow \infty} e^{- 2 \vert \alpha \vert ^2} {}_{1}F_{1}\left( \frac{1-p}{2},\frac{1}{2}, 2 \vert \alpha \vert^{2} \right) = 0, \quad p\ge 1. 
\end{eqnarray}

The actual value for the product of the quadratures standard deviation of our overcomplete, nonorthogonal basis states is given by the following expression,
\begin{eqnarray}
	(\sigma_{X}\sigma_{Y})^{2} &=& \left\{ \begin{array}{ll} 
	1 + 2 \vert \alpha \vert^{2} \Big[f(\vert \alpha \vert) - h^{2}(2, \vert \alpha \vert)\Big] - \frac{1}{4}\left( {\alpha^{\ast}}^{2} - \alpha^{2}\right)^{2} \Big[f(\vert \alpha \vert) - h^{2}(2, \vert \alpha \vert)\Big]^{2}, & p = 2, \\
	\frac{p^{2}}{4} + p \vert\alpha \vert^{2} \xi(p,\vert\alpha\vert) - \frac{1}{4}\left({\alpha^{\ast}}^{2} - \alpha^{2}\right)^{2} \xi^{2}(p,\vert\alpha\vert), & p \neq  2.
	\end{array} \right. \nonumber \\
\end{eqnarray}
where the auxiliary function is related to those defined before,
\begin{eqnarray}
\xi (p,\vert\alpha\vert)=\frac{(p-1)}{2\vert \alpha \vert^{2} (p-2)}g(p,\vert\alpha\vert) -h^{2}(p,\vert\alpha\vert)+1.
\end{eqnarray}
We can compare the lower bound of the Robertson uncertainty relation and the actual value for our para-Bose states and realize that they never minimize the uncertainty relation for the para-Bose field quadratures; see Figs. \ref{fig:Fig5}(a) and \ref{fig:Fig5}(d).

\section{Conclusion}

We have introduced an overcomplete, non-orthogonal basis for para-Bose Hilbert spaces of order $p$ using a structure that looks like standard boson Gilmore-Perelomov coherent states.
These states, for even para-Bose order, arise naturally in trapped ions where two orthogonal vibrational modes of the ion center of mass are involved.

Our overcomplete, nonorthogonal para-Bose basis shows the three types of Poissonian statistics in their mean excitation number distribution; this feature can be controlled by the para-Bose order and the complex parameter modulus. In an actual ion-trap experiment, these could be fine tuned as they are related to the initial state and the evolution time, in that order. 
Furthermore, our states present relative quadrature squeezing and never minimize the uncertainty relation for the para-Bose field quadratures standard deviations.

These properties could be used in a trapped-ion quantum simulation to generate tripartite states that inherit these nonclassical or semiclassical characteristics.

	\section*{Funding Information}
	The authors acknowledge financial support from CONACYT through doctoral Grants No. $\#455378$, No. $\#294880$, and project Grant No. CB-2015-01/$255230$.

	
\newpage	
\appendix
\section{Full extended forms}

The second momenta of the para-Bose excitation number,
	\begin{eqnarray} \nonumber
	\langle \hat{n}^2 \rangle &=& \left\{ \begin{array}{ll} 
	4 \vert \alpha \vert^{2} + \vert \alpha \vert^{4} + \frac{1}{\sqrt{2}}\vert \alpha \vert F\left(\sqrt{2}\vert \alpha\vert \right)-\frac{3}{2}\vert \alpha \vert^{2} f(\vert \alpha \vert) + \frac{2}{3}\vert \alpha \vert^{4} ~_{2}F_{2} \left(1,1; \frac{5}{2},3; -2 \vert \alpha \vert^2\right), & p = 2, \\ 
	\vert \alpha \vert^{2} + \vert \alpha \vert^{4} -3\sqrt{2}\Big[\vert \alpha \vert +\vert \alpha \vert^{3}\Big] F\left(\sqrt{2}\vert \alpha\vert \right) + \frac{9}{2} \vert \alpha \vert^{2} f(\vert \alpha\vert), & p = 4, \\
	p \vert\alpha\vert^{2} + \vert\alpha \vert^{4} + \frac{p-1}{6(p-4)(p-2)}\Big\{3\Big[ p - 1 - 2 \vert \alpha \vert^{2} (p-4)^{2} \Big]-\chi_{1}(p,\vert \alpha \vert) -\chi_{2}(p,\vert \alpha \vert)\Big\},& p \neq 2,4,
\end{array} \right.  \\ 
\end{eqnarray}
can be written in terms of Dawson integral $F(x)$ \cite{Olver2010} and the auxiliary integer functions,
\begin{eqnarray}
\chi_{1} (p,\vert\alpha\vert) &=& 3 e^{- 2 \vert\alpha\vert^2} \Big[p-1 + 2\vert \alpha \vert^{2} (5 p - 14) \Big]  ~_{1}F_{1}\left(\frac{3-p}{2}, \frac{3}{2}, 2\vert \alpha \vert^{2} \right) \\
\chi_{2}(p,\vert\alpha\vert) &=& 4 \vert \alpha \vert^{2} e^{- 2 \vert\alpha\vert^2 } (p-3)  \Big[1 - p + 
2\vert \alpha \vert^{2} (p + 2) \Big]  ~_{1}F_{1} \left( \frac{5-p}{2}, \frac{5}{2}, 2\vert \alpha \vert^{2} \right).
\end{eqnarray}
Its asymptotic expression for a large complex parameter modulus can be given in the following form,
	\begin{eqnarray} \nonumber
\langle \hat{n}^2 \rangle &\approx& \left\{ \begin{array}{ll} 
\vert \alpha \vert^{4} + \vert \alpha \vert^{2} \left[\ln \left(2 \vert \alpha \vert^{2}\right) + \frac{3}{2} - \psi \left(\frac{3}{2}\right)\right]+ \frac{1}{8} \ln \left(2 \vert \alpha \vert^{2}\right) - \frac{1}{4} + \frac{3}{8} \psi \left(\frac{1}{2}\right) - \frac{1}{2}\psi \left(\frac{3}{2}\right), & p = 2, \\
\frac{5 \vert \alpha \vert^{2}}{2} + \frac{9}{8} \left[ \ln \left(2 \vert \alpha \vert^{2}\right) - \psi \left(\frac{1}{2}\right)\right] - \frac{33}{16} , & p = 4, \\
p \vert\alpha\vert^{2} + \vert\alpha \vert^{4} + \frac{3(p-1)\big[ p - 1 - 2 \vert \alpha \vert^{2} (p-4)^{2} \big]}{6(p-4)(p-2)},& p \neq 2,4,
\end{array} \right. \vert \alpha \vert \gg 1. \\ 
\end{eqnarray}
With the full forms, we can calculate the standard deviation of the mean para-Bose excitation number, 
\begin{eqnarray} \nonumber
	\sigma_{n}^2 &=& \left\{ \begin{array}{ll} 
	4\vert \alpha \vert^{2} + \vert \alpha \vert^{4} + \frac{1}{\sqrt{2}}\vert \alpha \vert F\left(\sqrt{2}\vert \alpha \vert\right) - \frac{3}{2}\vert \alpha \vert^{2} f(\vert \alpha \vert) - \vert \alpha \vert^4 f^{2}(\vert \alpha \vert) +  \frac{2}{3}~_{2}F_{2}\left(1,1; \frac{5}{2},3; -2\vert \alpha \vert^{2}\right), p = 2, \\ 
	\vert \alpha \vert^{2} - 3\sqrt{2}\Big[\vert \alpha \vert +\vert \alpha \vert^{3}\Big] F\left(\sqrt{2}\vert \alpha\vert \right) + \frac{9}{2} \vert \alpha \vert^{2} f(\vert \alpha\vert) - \frac{9}{2} \vert \alpha \vert^{2} f^{2}(\vert \alpha\vert), \hspace{4cm} p = 4, \\
	\vert \alpha \vert^{2}\frac{3p-4}{p-2}-\frac{p-1}{12 (p-4) (p-2)^2}\Big(\Big\{3(p-4) (p-1)g^{2}(p,\alpha) + 2 (p-2) [\chi_{1}(p,\vert\alpha\vert) + \chi_{2} (p,\vert\alpha\vert) - 3 p+3]\Big\} & \\
	+ 12\vert \alpha \vert^{2} (p-4) (p-2)g(p,\vert\alpha\vert)\Big) , \hspace{9cm}  p \neq 2,4
	\end{array} \right.  \\ 
	\end{eqnarray}

These full forms yield the following Mandel Q parameter,
	\begin{eqnarray} \nonumber
	Q &=& \left\{ \begin{array}{ll} 
	\frac{1}{ 6\vert \alpha \vert f(\vert\alpha\vert)} \Big[4 \vert \alpha \vert^3 ~_{2}F_{2}\left(1,1; \frac{5}{2},3; -2 \vert \alpha \vert^2\right) + 6 \vert \alpha \vert^3 + 24 \vert \alpha \vert + 3 \sqrt{2}F\left(\sqrt{2} \vert \alpha \vert \right)\Big] -\frac{5}{2} - \vert \alpha \vert^2 f(\vert\alpha\vert) ,\; p = 2, \\ 
	\frac{3}{2 \vert \alpha \vert + 3 \sqrt{2} F\left(\sqrt{2} \vert \alpha \vert \right)} \Big\{3 \vert \alpha \vert  \Big[ f(\vert\alpha\vert) - F^{2}\left(\sqrt{2} \vert \alpha \vert \right) \Big] - \sqrt{2} \left(4 \vert \alpha \vert^2 + 3\right) F\left(\sqrt{2} \vert \alpha \vert \right)\Big\} , \hspace{2.5cm} \; p = 4, \\
	\frac{ (p-1)}{6 (p-4) (p-2) \Big[2\vert \alpha \vert^2 (p-2) + (p-1) g(p,\vert\alpha\vert)  \Big]} \Big\{ - 6\left(2 \vert \alpha \vert^2 + 1\right) (p-4) (p-2) g(p,\vert\alpha\vert) \Bigg. &\\
	\Bigg. -3(p-4) (p-1) g^{2}(p,\vert\alpha\vert) - 2 (p-2) [\chi_{1}(p,\vert\alpha\vert) + \chi_{2}(p,\vert\alpha\vert) - 12\vert\alpha\vert^2 (p-4) -3p + 3 ]\Big\}, p \neq 2,4.
	\end{array} \right. \\ 
	\end{eqnarray}

The asymptotic values of the para-Bose field quadratures for large complex parameter are:
\begin{eqnarray}
\langle \hat{X} \rangle &\approx& \frac{1}{\sqrt{2}} \left(\alpha + \alpha^{\ast}\right), \nonumber \\
\langle \hat{Y} \rangle &\approx& \frac{i}{\sqrt{2}} \left(\alpha^{\ast} -  \alpha\right), \qquad \vert\alpha\vert \gg 1. 
\end{eqnarray}
The exact full form of the second momenta for the field quadratures is given in the following,
\begin{eqnarray} \nonumber
\langle \hat{X}^2 \rangle &=& \left\{ \begin{array}{ll} 
1 + \frac{1}{2}\left(\alpha + \alpha^{\ast}\right)^{2} f(\vert\alpha\vert) , & p = 2, \\ 
\frac{p}{2} + \left(\alpha + \alpha^{\ast}\right)^{2} \left[\frac{1}{2} + \frac{(p-1)}{4(p-2)\vert \alpha \vert^{2}} g(p,\vert\alpha\vert) \right] , & p \neq 2,
\end{array} \right. \nonumber \\
\langle \hat{Y}^2 \rangle &=& \left\{ \begin{array}{ll} 
1 - \frac{1}{2}\left( \alpha^{\ast} -\alpha \right)^{2} f(\vert\alpha\vert) , & p = 2, \\ 
\frac{p}{2} - \left(\alpha^{\ast} - \alpha\right)^{2} \left[\frac{1}{2} + \frac{(p-1)}{4(p-2)\vert \alpha \vert^{2}} g(p,\vert\alpha\vert) \right], & p \neq 2,
\end{array} \right. \nonumber \\ 
\end{eqnarray}
and has an asymptotic approximation,
\begin{eqnarray} \nonumber
\langle \hat{X}^2 \rangle &\approx& \left\{ \begin{array}{ll} 
1 + \frac{1}{2}\left(\alpha + \alpha^{\ast}\right)^{2}\left\{1 + \frac{1}{4 \vert \alpha \vert^{2}}\left[\ln \left(2 \vert \alpha \vert^{2}\right)-\psi \left(\frac{1}{2}\right)\right]\right\}, & p = 2, \\ 
\frac{p}{2} + \left(\alpha + \alpha^{\ast}\right)^{2} \left[\frac{1}{2} + \frac{(p-1)}{4(p-2)\vert \alpha \vert^{2}} \right] , & p \neq 2,
\end{array} \right. \nonumber \\
\langle \hat{Y}^2 \rangle &\approx& \left\{ \begin{array}{ll} 
1 - \frac{1}{2}\left( \alpha^{\ast} -\alpha \right)^{2}\left\{1 + \frac{1}{4 \vert \alpha \vert^{2}}\left[\ln \left(2 \vert \alpha \vert^{2}\right)-\psi \left(\frac{1}{2}\right)\right]\right\} , & p = 2, \\ 
\frac{p}{2} - \left(\alpha^{\ast} - \alpha\right)^{2} \left[\frac{1}{2} + \frac{(p-1)}{4(p-2)\vert \alpha \vert^{2}} \right], & p \neq 2,
\end{array} \right. \vert\alpha\vert \gg 1. \nonumber \\ 
\end{eqnarray}
The full form for their variances can also be given in the asymptotic limit,
\begin{eqnarray}
\sigma^{2}_{X} &\approx& \left\{ \begin{array}{ll} 
1 + \frac{1}{2} \left(\alpha + \alpha^{\ast}\right)^{2} \Big[\frac{1}{4 \vert\alpha \vert^2} \left[\ln \left(2 \vert \alpha \vert^2\right)-\psi \left(\frac{1}{2}\right)\right] - \frac{1}{2 \sqrt{\pi } \vert \alpha \vert^2} \left(2 + \frac{1}{2 \sqrt{\pi } \vert\alpha \vert^2} \right)\Big] , & p = 2,  \\
\frac{p}{2} + \frac{1}{2}\left(\alpha + \alpha^{\ast}\right)^{2} \left[\frac{(p-1)}{2 \vert \alpha \vert^{2} (p-2)}\right], & p \neq 2,
\end{array} \right. \nonumber \\  
\sigma^{2}_{Y} &\approx& \left\{ \begin{array}{ll} 
1 - \frac{1}{2} \left(\alpha^{\ast} - \alpha\right)^{2} \Big[\frac{1}{4 \vert \alpha \vert^2} \left[\ln \left(2 \vert \alpha \vert^2\right)-\psi \left(\frac{1}{2}\right)\right] - \frac{1}{2 \sqrt{\pi } \vert \alpha \vert^2} \left(2 + \frac{1}{2 \sqrt{\pi } \vert\alpha \vert^2} \right)\Big] , & p = 2,  \\
\frac{p}{2} - \frac{1}{2}\left(\alpha^{\ast} - \alpha\right)^{2} \left[\frac{(p-1)}{2 \vert \alpha \vert^{2} (p-2)}\right], & p \neq 2,
\end{array} \right.  \vert \alpha \vert \gg 1.\nonumber \\
\end{eqnarray}
Finally, this is the full form for the product of the quadratures variances in the asymptotic limit,
\begin{eqnarray}
	(\sigma_{X}\sigma_{Y})^{2} &\approx& \left\{ \begin{array}{ll} 
		1 + \frac{1}{2} \left[\ln \left(2 \vert \alpha \vert^2\right)-\psi \left(\frac{1}{2}\right)\right] - \frac{1}{ \sqrt{\pi }} \left(2 + \frac{1}{2 \sqrt{\pi } \vert\alpha \vert^2} \right) & \\
		\quad + \frac{1}{4}\left(\alpha^{2} + {\alpha^{\ast}}^{2}\right)^{2} \Big[\frac{1}{4 \vert\alpha \vert^2} \left[\ln \left(2 \vert \alpha \vert^2\right)-\psi \left(\frac{1}{2}\right)\right] - \frac{1}{2 \sqrt{\pi } \vert \alpha \vert^2} \left(2 + \frac{1}{2 \sqrt{\pi } \vert\alpha \vert^2} \right)\Big]^2, & p = 2, \\
		\frac{p^{2}}{4} + \frac{p(p-1)}{2(p-2)} + \frac{1}{4} \left(\alpha^{2} + {\alpha^{\ast}}^{2}\right)^{2} \left[\frac{(p-1)}{2(p-2)\vert \alpha \vert^{2}}\right]^{2} , & p \neq  2.
	\end{array} \right. \nonumber \\
\end{eqnarray}


%

\end{document}